\documentclass[11pt]{article}
\usepackage{titling}

\usepackage{times}

\usepackage{amsmath}
\usepackage{amsfonts}
\usepackage{amssymb}
\usepackage{graphicx}
\usepackage{color}

 \newcommand{\EF}{$\mathrm{E}_\mathrm{F}$}

\newcommand{\hn}{$\mathrm{h\nu}$}

 \newcommand{\kpar}{$k_\parallel$\ }
 \newcommand{\kperp}{\hbox{$k_\perp$}}

 \newcommand{\Gbar}{$\overline{\Gamma}$}

 \newcommand{\KGK}{$\overline{\mathrm{K}\Gamma\mathrm{K}}$}
 
\newcommand{\ZU}{$\mathrm{Z}$-$\mathrm{U}$}
\newcommand{\ZA}{$\mathrm{Z}$-$\mathrm{A}$}

 \newcommand{\AZA}{$\mathrm{A}$-$\mathrm{Z}$-$\mathrm{A}$}

\newcommand{\aaGeTe}{$\alpha$\nh GeTe}
\newcommand{\GMT}{Ge$_{1-x}$Mn$_x$Te}
\newcommand{\GMTD}{Ge$_{0.87}$Mn$_{0.13}$Te}
\newcommand{\GMTDD}{Ge$_{0.97}$Mn$_{0.03}$Te}

\newcommand{\n}[1]{$n$\nobreakdash-\hspace{0pt}}
\newcommand\nh{\mbox{-}}
\newcommand\degC{$\,^{\circ}\mathrm{C}$}
\newcommand\dg{$^{\circ}$}

\newcommand{\Pz}{$P_{z}$}

\title{Entanglement and manipulation of the magnetic and spin-orbit order in multiferroic Rashba semiconductors.} 
\author{J.~Krempask{\'y}${}^{1}$, S.~Muff${}^{1,2}$,  F.~Bisti${}^{1,}$,  M.~Fanciulli${}^{1,2}$, H.~Volfov{\'a}${}^{3}$, A.~Weber${}^{1,2}$, \\ N.~Pilet${}^{1}$,  P.~Warnicke${}^{1}$,  H.~Ebert${}^{3}$, J.~Braun${}^{3}$, F.~Bertran${}^{4}$, V.V.~Volobuev${}^{5,7}$,  \\
J.~Min{\'a}r${}^{3,6}$, G.~Springholz${}^{7}$, J.~H.~Dil${}^{1,2}$, V.N.~Strocov${}^{1}$\\
\\
\normalsize{${}^{1}$Swiss Light Source, Paul Scherrer Institut, CH-5232 Villigen PSI, Switzerland}\\
\normalsize{${}^{2}$Institute of Physics, \'Ecole Polytechnique F\'ed\'erale de Lausanne, CH-$1015$ Lausanne, Switzerland}\\
\normalsize{${}^{3}$Department of Chemistry, Ludwig Maximillian University, 81377 Munich, Germany}\\
\normalsize{${}^{4}$SOLEIL Synchrotron, L'Orme des Merisiers, F-91192 Gif-sur-Yvette, France}\\
\normalsize{${}^{5}$National Technical University, Kharkiv Polytechnic Institute, Frunze Str. 21, 61002 Kharkiv, Ukraine}\\
\normalsize{${}^{6}$New Technologies-Research Center University of West Bohemia, Plze{\v n}, Czech Republic}\\
\normalsize{${}^{7}$Institut f{\"u}r Halbleiter-und Festk{\"o}rperphysik, Johannes Kepler Universit{\"a}t, A-4040 Linz, Austria}\\
\\
}

\date{}

\begin{document} 
\maketitle 
\vspace*{-1.8cm}
\begin{abstract}
The interplay between electronic eigenstates, spin, and orbital degrees of freedom, combined with fundamental breaking of symmetries is currently one of the most exciting fields of research. It not only forms the basis for giant magnetoresistance \cite{Fert}, spin-torque manipulation of magnetic domains \cite{Miron:2010} and the use of Rashba effects for spin manipulation \cite{Bychkov:1984}, but has also lead to outstanding recent discoveries of new quantum phases such as topological insulators \cite{Hasan:2010}, Weyl semimetals \cite{Murakami:2007} and Majorana fermions \cite{Mourik:2012}. While already each of these topics has initiated new research fields, the combination of several of these effects in a single material opens up a new realm of opportunities. For example, superconductors with sizable Rashba spin splitting and ferromagnetic order would unite all ingredients for formation and manipulation of so-called anyons, generalizing the concept of Majorana fermions \cite{Beenakker:2013}. Multiferroics, such as \GMT\ \cite{Springholz_PRL}, fulfill these requirements providing unusual physical properties \cite{Nat_Eerenstein, Nat_Ramesh, ADMA_Tokura} due to the coexistence and coupling between ferromagnetic and ferroelectric order in one and the same system. Here we show that multiferroic \GMT\  inherits from its parent ferroelectric \aaGeTe\ compound a giant Rashba splitting of three-dimensional bulk states  \cite{Picozzi_AdvM,JK_GeTe} which competes with the Zeeman spin splitting induced by the magnetic exchange interactions. The collinear alignment of ferroelectric and ferromagnetic polarization leads to an opening of a tunable Zeeman gap of up to 100 meV around the Dirac point of the Rashba bands, coupled with a change in spin texture by entanglement of magnetic and spin-orbit order. Through applications of magnetic fields, we demonstrate manipulation of spin-texture by spin resolved photoemission experiments, which is also expected for electric fields based on the multiferroic coupling. The control of spin helicity of the bands and its locking to ferromagnetic and ferroelectric order opens fascinating new avenues for highly multifunctional multiferroic Rashba devices suited for reprogrammable logic and/or nonvolatile memory applications.
\end{abstract}

\begin{figure}[ht!]
\centering
\includegraphics[width=15cm]{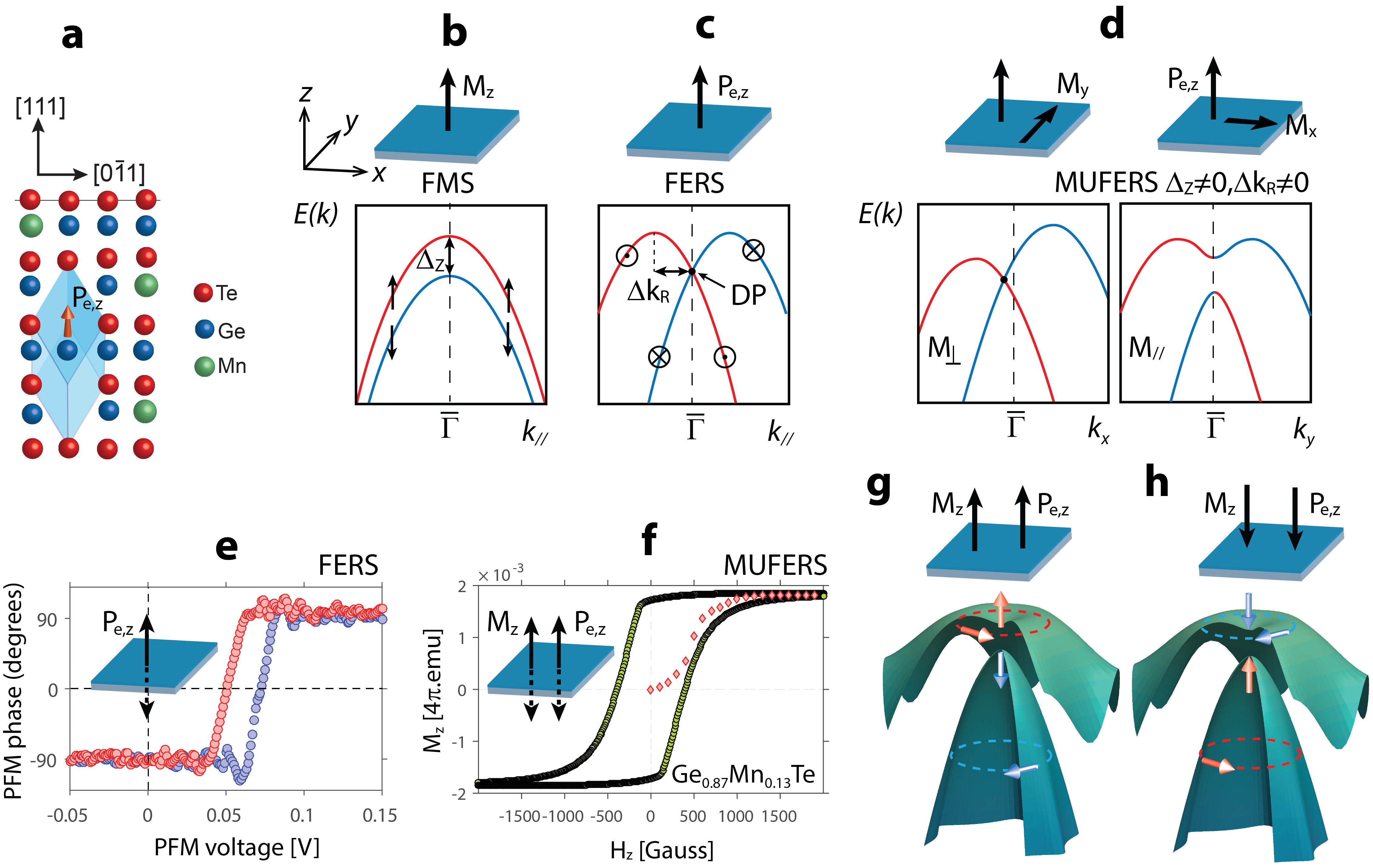}
\caption{\textbf{Basic properties.} \textbf{(a)} Sketch of  multiferroic \GMT\  with ferroelectric displacement of Ge(Mn)-atoms inside the rhombohedrally distorted unit cell along [111] as indicated by the orange arrow. \textbf{(b)} Schematic Rashba-gas band maps of a ferromagnetic semiconductor (FMS) compared to a ferromagnetic Rashba semiconductor (FERS) \textbf{(c)} and a multiferroic Rashba semiconductor (MUFERS) in \textbf{(d)}, with their dependence on the orientations of the ferromagnetic ($M$) and ferroelectric ($P_e$) order (see text). \textbf{(e)} Typical out-of-plane ferroelectric phase hysteresis measured by piezo-force-microscopy for ferroelectric GeTe and preserved in \GMT\  \cite{Springholz_PRL}. \textbf{(f)} Out-of-plane ferromagnetic hysteresis curve of multiferroic \GMTD\  measured by SQUID. \textbf{(g-h)} Band plots of the two upper occupied RZ-split valence bands as a function of k-vectors around Z-point of \GMT\ and corresponding spin-texture switching upon colinear reversal of $M_z$ and $P_{z}$.} 
\label{F0}
\end{figure}

\noindent Similar to how the magnetic moment of an unpaired electron strongly enhances the exchange interaction, the local spin-orbit interaction is greatly enhanced by broken inversion symmetry. The larger the symmetry breaking, the larger the Rashba spin splitting becomes \cite{Bihlmayer:2006,Dil:2009R,Slomski:2013}. This is exemplified by the giant spin splitting observed for the polar semiconductor BiTeI \cite{BiTeI_NC,BiTeI_Hugo}, the polar surface of SrTiO$_3$  \cite{Santander:2014}, as well as for high-Z surface alloys \cite{Ast:2007,Gierz:2010}. In ferroelectric materials, the inversion symmetry is naturally broken by the ferroelectric (FE) polarization produced by the relative anion/cation  displacements in the crystal lattice. Thus, the spin splitting in ferroelectric Rashba semiconductors (FERS) can assume exceedingly high values.  \aaGeTe\ exhibits record splitting values \cite{Picozzi_AdvM,JK_GeTe} due to the very large rhomobohedral lattice distortion providing cation/anion displacements  as large as 10\% of the (111) lattice plane spacing (Fig.~1a). The spin splitting of its bulk bands is intimately linked to the FE polarization, meaning that it can be switched and controlled by reversing the ferroelectric polarization direction \cite{Picozzi_AdvM}.
Doping of GeTe with Mn, \GMT\ maintains the rhombohedral lattice distortion and FE properties up to Mn concentrations as high as 30\% \cite{Springholz_PRL}. At the same time, the Mn spins couple to each other via  free carrier mediated RKKY exchange interactions. As a result, ferromagnetic (FM) ordering as in (GaMn)As \cite{Masaki_GaMnAs} occurs, which turns \GMT\ into a multiferroic Rashba semiconductor (MUFERS) that combines both ferroelectric and ferromagnetic properties as shown by Fig.~1e and f, respectively.  In fact, multiferroic \GMT\ features one of the highest FM Curie temperatures \cite{Springholz_PRL,BaF2_1} and highest reported magnetic moments amongst diluted ferromagnetic semiconductors (FMS) \cite{APL_Fukuma}.

Ferromagnetic ordering and ferroelectric symmetry breaking both lead to a spin splittings of the electronic band structure. For FMS, the spontaneous magnetization leads to a vertical Zeeman splitting of the bands (Fig.~1b), whereas for FERS the electric polarization induces a Rashba splitting in the \kpar\-direction (Fig.~1c). 
In MUFERS both effects are entangled with each other, modifying the spin texture and band dispersions (Fig. 1d). For a basic discussion, we consider a simplified Rashba-Zeeman model based on a two-dimensional free-electron approximation (Rashba-Zeeman gas) in which the energy eigenvalues $E_\pm(k)$ of the Rashba-split bands in the presence of magnetic order are given by:
\begin{equation*}
  E_\pm(k) = E_0 + \frac{\hbar^2k^2}{2m^*} \pm\sqrt{\Delta^2_Z + 4\alpha^2k^2},
 \label{1}
\end{equation*}
where $E_0$ is the band edge, $m^*$ the effective electron mass, $\alpha$ the $k$-linear Rashba coupling constant and $\Delta_Z$ the momentum-independent Zeeman gap near \Gbar. In principle, any orientation of the magnetization opens a gap $\Delta_Z$ near \Gbar\ (see Fig.~1d). However, considering the momentum-dependent expected values of the spin components in all directions (see Supplemental Note 1), only systems with magnetization parallel to the inversion symmetry breaking direction can reorient the spin of the Rashba electron and open a gap in all \kpar directions as illustrated by Fig.~1g. Due to coupling between the easy axis of magnetization and the rhombohedral lattice distortion, \GMT\ (111) films fulfill these criteria \cite{Springholz_PRL} because the electric polarization and magnetic moments are colinear and perpendicular to the surface (Fig.~1e,f). Since the helicity of the Rashba effect is locked to the ferroelectric moments \cite{JK_GeTe,Liebmann_GeTe,Schoenhense_GeTe}, multiferroic switching is expected to entangle with the spin helicity as sketched in Fig.~1g-h.

\begin{figure}[ht!]
\centering
\includegraphics[width=14cm]{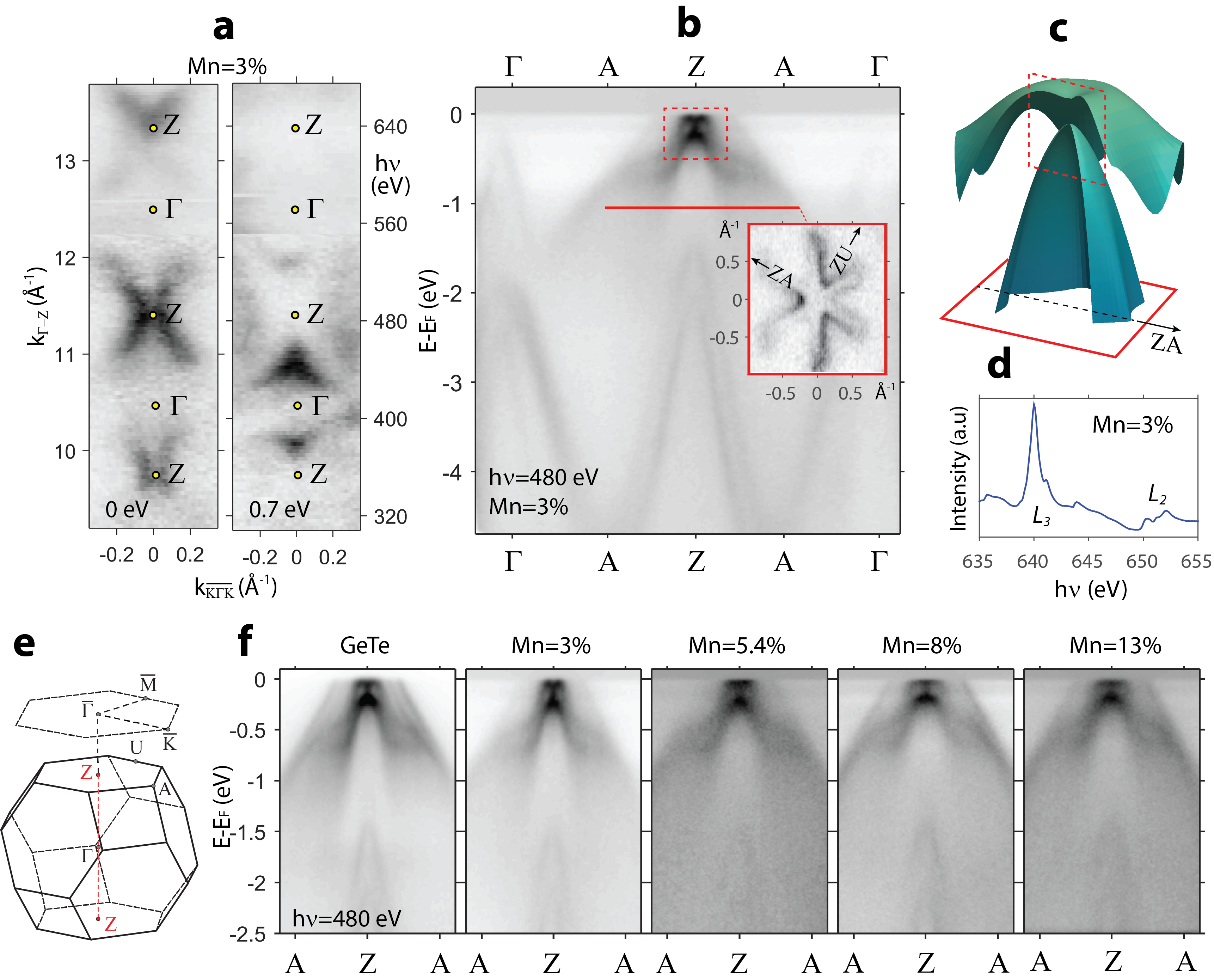}
\caption{\textbf{Soft x-ray ARPES data.}  \textbf{(a)} SXARPES photon-energy dependent constant binding-energy maps of the \GMTDD\ valence band bulk states near Fermi level and at 0.7 eV binding energy in the \KGK\ plane.  \textbf{(b)}  SXARPES band map measured along \hbox{$\mathrm{Z}$-$\mathrm{A}$-$\Gamma$} with theoretical band plots in \textbf{(c)} to compare with measured data near \EF. The inset is a constant binding energy map at 1 eV binding energy. \textbf{(d)} X-ray absorption data near the $L_3$ absorption edge. \textbf{(e)} Brillouin zone of quasicubic GeTe. \textbf{(f)} ARPES band maps along \AZA\ for the selected Mn-dopings. The dashed rectangle in \textbf{(b)} zoom into region near \EF\ where the Zeeman gap opening is examined.}
\label{F1}
\end{figure}

For a systematic spectroscopic study of the Rashba-Zeeman (RZ) splitting in this multiferroic system, we have prepared epitaxial \GMT\ films with various Mn\nh content up to 13\% ($x_{Mn}$ 0.03, 0.054, 0.08 and 0.13, see Methods). We used soft \hbox{X-ray} ARPES (SX-ARPES) to explore the local electronic structure of the Mn\nh ions and RZ\nh splitting with three-dimensional (3D) \textbf{k}-space resolution, and spin-resolved ARPES (SARPES) in the UV photon energy range to explore the corresponding changes in the spin textures in this multiferroic material (see Methods). Fig.~2 summarizes our SX-ARPES data obtained with photon energies from 340 to 800 eV. A fundamental advantage of this energy range  \cite{Strocov:2014} is an increase of the photoelectron escape depth and thus of intrinsic definition of the ARPES experiment in surface-perpendicular momentum \kperp\ \cite{Strocov_kz} crucial for observation of the inherently 3D electronic structure of \GMT. This is illustrated in Fig.~2a by showing the dispersive spectral weight at the Fermi level (\EF) and at a binding energy of 700 meV as a function of \kperp\ varied through photon energy. This map, that is similar for all samples except for an increase of the spectral broadening with Mn concentrations $x$, readily identifies the Z-points in the 3D Brillouin zone where the Rashba splitting is most pronounced \cite{JK_GeTe}. Measured in the Z-point at $h\nu=480$ eV, where the spectral intensity is maximal, the ARPES spectral weight along the \AZA\ direction is shown in Fig.~2b. This map clearly shows the pair of Rashba-split bands in the vicinity of the Z-point consistent with the theoretical expectation (Fig.~2c). Another virtue of SX-ARPES is its chemical specificity achieved with resonant photoemission. The corresponding data measured at the Mn \textit{L}-edge as indicated by the X-ray absorption (Fig.~2d) demonstrates that the Mn 3\textit{d} states hybridize into the GeTe host states through the whole valence band, in particular into the Rashba bands below the Zeeman gap (Supplemental Note 2). In particular, the absence of any impurity states in the vicinity of \EF\ which might otherwise interfere with the RZ\nh splitting in this region.

\begin{figure}[b!]
\centering
\includegraphics[width=15cm]{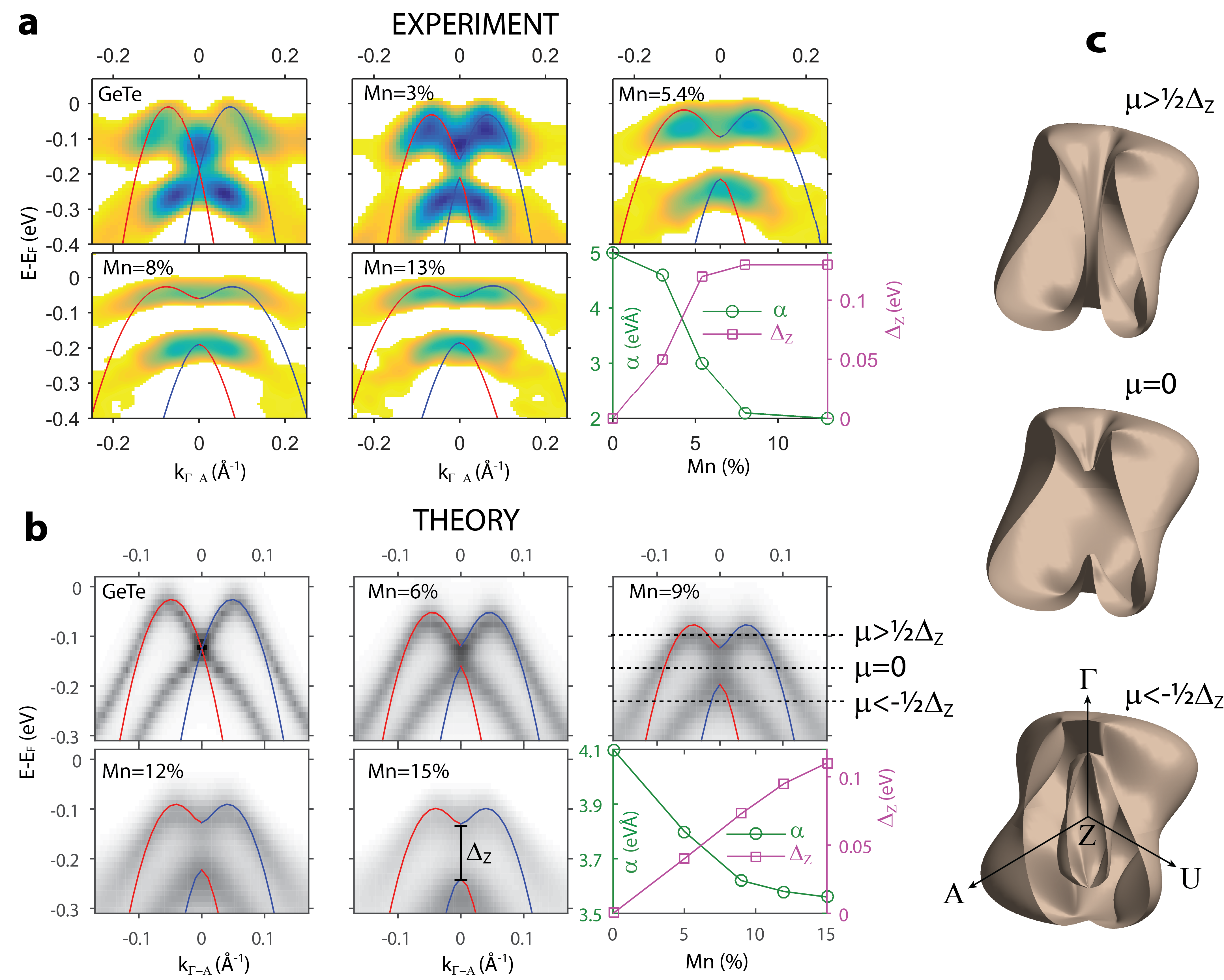}
\caption{\textbf{Rashba-Zeeman splitting: experiment, theory and consequences for 3D Fermi surface.}  \textbf{(a)} Zoomed-in ARPES data from Fig.~2f (second derivative plots from red rectangle area in Fig.~2b) compared to \textit{ab initio}  \GMT\ calculations for selected Mn\nh dopings in \textbf{(b)}; each fitted to a simplified Rashba-Zeeman gas model (red/blue lines). The best-fit parameters of the Rashba parameter $\alpha$ and Zeeman gap opening $\Delta_Z$ are displayed in the corresponding lower right hand panels. \textbf{(c)} 3D Fermi surfaces for three different chemical potentials $\mu$.}
\label{F2}
\end{figure}

Our further discussions of the Zeeman gap opening in \GMT\ is based on Fig. 2f, which presents  the experimental band structure along A-Z-A in vicinity of the Z-point, measured through the whole sample series. To elucidate in detail the correlation between Rashba splitting and Zeeman gap opening, and its dependence on Mn doping, we compare in Fig.~3a,b the zoom-in band maps around the Z-point near \EF\ (dashed rectangle in Fig.~2b) with \textit{ab initio} calculations. As detailed in Supplemental Note 3, careful experimental alignment is needed for the quantitative assessment of the Zeeman gap. To highlight the competing Rashba and Zeeman terms $\alpha$ and $\Delta_Z$ for different doping levels, both theoretical and experimental data are fitted with the simplified Rashba-Zeeman gas model (solid lines) described above.
Due to the simplicity of the model that neglects the strong band non-parabolicity of the IV-VI compounds, the fit does not reproduce the actual dispersions but it gives direct values for both $\alpha$ and $\Delta_Z$. 
We observe both in experiment and theory that a Zeeman gap, absent in \aaGeTe, appears and widens with increasing Mn concentration, i.e. increasing sample magnetization. This is clear-cut evidence that the Zeeman gap opening in the Dirac point of the Rashba bands is induced by the ferromagnetic order of the multiferroic system. 
To the best of our knowledge this is the first experimental confirmation of the opening of a Zeeman gap at the Dirac point in a system with strong ferromagnetic order, which was so far elusive in magnetically doped topological insulators due to the lack of measurements justifying the ferromagnetic ordering of the dilute dopants at the ARPES measurement conditions \cite{MDTI_ZX_Science,MDTI_Hasan_NPhys, Rader:2016}.
Plotting in  the lower right panels of Fig.~3a,b the resulting values for $\alpha$ and $\Delta_Z$ as a function of Mn content reveals that the Zeeman gap reaches remarkably high values around 100 meV for $x_{Mn}>$ 10\%. 
Furthermore, we observe that the progression of the Zeeman gap is accompanied by a decrease of the Rashba splitting. The microscopic origin of this trend is that higher Mn-doping reduces the rhombohedral distortion of the \GMT\ lattice \cite{Springholz_PRL}. This, in turn, reduces the non-centrosymmetric arrangement of the Ge cations  and FE polarization (cf. Fig.~1a)  and thus, the strength of the Rashba coupling constant.

A particularly interesting feature for device applications is the fact that with increasing Mn doping the Zeeman gap shifts towards \EF. This will reduce the bias voltage needed to move \EF\ inside this gap and create a single spin polarised Fermi surface reminiscent of an ideal topological insulator \cite{Hasan:2010}.  In contrast to the 2D surface state of a topological insulator, however, \GMT\ has a three dimensional Fermi surface, which in the absence of ferromagnetic order takes the form of a spin polarized spindle torus \cite{BiTeI_Hugo,Landolt:2015}.  The constant energy surface shows a strong hexagonal warping which becomes three-fold away from the $\Gamma$ and $\mathrm{Z}$ points, in accordance with the crystal symmetry. In Fig.~3c the 3D Fermi surfaces of \GMT\ are shown for three different chemical potentials, taking into account both the deviation from an ideal torus due to the warping and the opening of a Zeeman gap. For $\mu>\frac{1}{2}\Delta_Z$ the gap opening does not influence the Fermi surface, and for $\mu<-\frac{1}{2}\Delta_Z$ the gap is only visible for a limited \kperp\ range because of the dispersion to higher binding energy away from $\mathrm{Z}$. For $\mu=0$, however, the chemical potential is within the Zeeman gap and the 3D Fermi surface obtains a non trivial spin texture due to the presence of only one single non-degenerate Fermi sheet. The exact consequences of this configuration require further study, but in both the 1D and 2D equivalents this leads to a template for the formation of Majorana fermions.

We now turn to spin-resolved information achieved with  UV-SARPES to show that \GMT\ still retains the Rashba-type spin splitting from the host GeTe, but exhibits an out-of-plane spin reorientation due to ferromagnetic ordering. According to our previous studies, \aaGeTe\ already exhibits a complex surface electronic structure where surface states coexist with a variety of bulk derived states \cite{JK_GeTe,Liebmann_GeTe}. For \GMT, the data measuring the in-plane spin polarization shown in Supplemental Note 4.1 and 4.2 indicate opposite spin helicity of the bulk states above and below the Zeeman gap as sketched Fig.~1g. Thus, the in-plane spin polarization is similar as for \aaGeTe\ above and below the Dirac point \cite{JK_GeTe}.

\begin{figure}[b!]
\centering
\includegraphics[width=14cm]{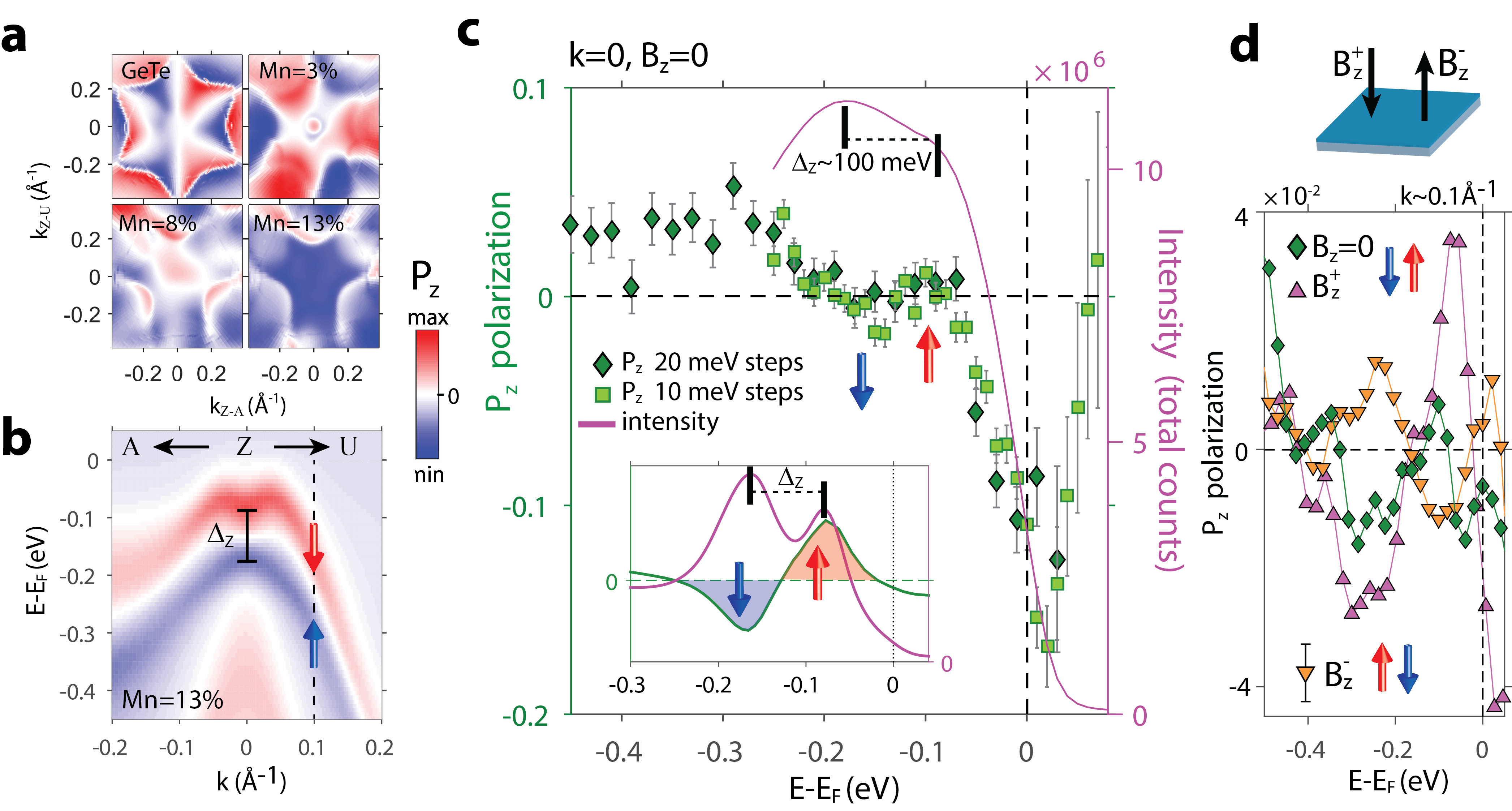}
\caption{\textbf{Intrinsic and magnetized out-of-plane spin polarization: theory and experiment}. \textbf{(a)} Calculated out-of-plane spin polarization in the photoemission final states for \hn=22 eV for GeTe below the Dirac point and for \GMT\ with $x_{Mn}$ of 3\%, 8\% and 13\% below the Zeeman gap ($\approx$0.2 eV binding energy). \textbf{(b)} \GMTD\ initial-state calculations based on multiple scattering alloy theory along \ZA\ and \ZU. \textbf{(c)} Corresponding measured SARPES \Pz-polarization in the Z-point visualized as energy-distribution curves in \Pz\ and total intensity measured across the Zeeman gap. The theoretical data from \textbf{(b)} is shown for comparison in the in-set. The data points correspond to  two independent experimental data-sets with 20 meV and 10 meV energy steps. The vertical arrows highlight the \Pz-wiggle resolved across the Zeeman gap $\Delta_Z$ of the order of 100 meV. \textbf{(d)} A comparison of \Pz-spin reorientation between intrinsic and magnetized \GMTD\ along the $[\bar{1} \bar{1} \bar{1}]$ ($B_{z}^+$) and $[111]$ ($B_{z}^-$) directions.}
\label{F3}
\end{figure}

To resolve the question of how the presence of ferromagnetic order affects the \GMT\ spin textures, Fig.~4a  presents \Pz\nh calculations for four different Mn concentrations based on the one-step photoemission approach (see Methods). Clearly, ferromagnetic ordering leads to a progressive \Pz\ spin reorientation with increasing Mn-concentration. In  particular, the characteristic \Pz\ hexagonal warping typical for GeTe is smeared out at \Gbar\  (Supplemental Note 4.3). We will now visualize the theoretical and experimental \Pz\ spinors as spin-resolved energy distribution curves measured across the Zeeman gap at normal emission. 
Our experimental data in Fig.~4c,d and the theoretical data represented by the Fig.~4c inset extracted from panel (b), consistently reproduce the characteristic \Pz\ spin reorientation "wiggle" across the Zeeman gap induced by spontaneous FM order.

The virtue of MUFERS obviously lies in the ample possibilites for spin manipulation using electrical as well as magnetic fields or by combination of both (see Fig. 1). The latter is demonstrated by Fig. 4(d), where we present and compare the \Pz\ spin texture measured before and after magnetization reversal, i.e., after switching the magnetization between the  $[111]$ and  $[\bar{1}\bar{1}\bar{1}]$ directions with magnetic fields $\pm B$  perpendicular to the sample. After application of $B$  along $[\bar{1}\bar{1}\bar{1}]$, the \Pz\nh polarization is significantly enhanced, meaning that the intrinsic FM order is into the sample. On the other hand, after switching the magnetization in the opposite $[111]$ direction, the \Pz\ spin-texture reverts, which clearly confirms the entanglement of the spin-splitting as sketched in Fig.~1g,h. In order to illustrate the \Pz-spin reorientation around \Gbar, the SARPES data in Fig.~4d was measured for finite momenta indicated in Fig.~4b. They show exactly the same \Pz\nh modulation as in normal emission, which is in sharp contrast to pure \aaGeTe\ where the out-of-plane spin texture is dominated by warping and photoemission effects \cite{JK_GeTe} (Supplemental Note 4.3).
As revealed by Fig. 1(f), for GeMnTe magnetic field strengths of less than 1000 Gauss are sufficient for this reversal process, where the actually required switching field depends on temperature and Mn concentration \cite{Springholz_PRL}.
\newline
\indent
In a broader perspective, our work introduces a new paradigm of multi-ferroic rashba semiconductors, which generalizes and  expands the concept of FERS and FMS. The unveiled Rashba-Zeeman splitting has far reaching consequences both for fundamental physics and device applications.  From a theoretical point of view the Zeeman gap of MUFERS may give rise to excitations with the characteristics of localized Majorana fermions. In terms of applications, adding magnetism to ferroelectric FERS  opens additional degrees of freedom and allows to strongly enhance the fidelity of spin control through additional Larmor precession of spins injected in field effect spin transistors. Spin control in such devices can be attained through electric or magnetic means, as well as by combinations of both. 
Combining the active channel and magnetic material in a single material without needing to worry about interfaces is relevant for spintronics technology. This  vastly enhances the functionality of MUFERS devices beyond that of ferromagnetic or ferroelectric systems only \cite{Rinaldi_2016}.
Thus, our results will pave the way for a new field of multiferroic Rashba semiconductor systems, bringing new multifunctional assets for spintronic device applications.

\section*{Methods}
\subsection*{Sample preparation method.}
Experiments were performed on 200 nm thick \GMT\ films grown by molecular beam epitaxy on BaF$_2$(111) substrates \cite{Springholz_PRL,BaF2_1,BaF2_2}. A protective stack of amorphous Te- and Se-capping layers with a total thickness of ~20 nm was used to avoid surface oxidation and degradation. It was  completely removed in the ultrahigh ARPES vacuum chamber by annealing the samples for 30-45 min at 250\degC. Comparison of SX-ARPES spectra measured through the protective Te/Se stack and on the uncapped samples evidenced that the annealing did not change the \GMT\ bulk electronic structure but only unleashed the otherwise suppressed surface states. 

\subsection*{Experimental techniques.}
The bulk-sensitive SX-ARPES experiments have been carried out at the SX-ARPES endstation \cite{Strocov_ES} of the high-brilliance ADvanced RESonance Spectroscopies (ADRESS) beamline at the Swiss Light Source synchrotron radiation facility, Paul Scherrer Institute, Switzerland, using $p$-polarized soft-X-ray photons in the energy range 300--800 eV. The slit of the hemispherical photoelectron analyzer PHOIBOS-150 was oriented in the scattering plane including the incident photons and detected photoelectrons. The combined (beamline and analyzer) energy resolution was ~80 meV, and analyzer angular resolution ~0.07\dg. The experiments were carried out at low temperature of 12 K to quench the thermal effects destroying the coherent k-resolved spectral component at high photon energies \cite{JBraun_XPSlimit}.  The complementary SARPES experiments were performed at the COPHEE end-station of the Surfaces and Interfaces Spectroscopy (SIS) beamline at SLS using $p$-polarized photons in the energy range 20--25 eV. The Omicron EA 125 hemispherical energy analyzer was equipped with two orthogonally mounted classical Mott detectors\cite{Hoesch_JESRP}. The whole set-up allows simultaneous measurements of all three spatial components of the spin-polarization vector for each point of the band structure. The SARPES data were measured with the sample azimuths \ZU\ or \ZA\ aligned perpendicular to the scattering plane. The angular and combined energy resolution were 1\dg~and 60 meV, respectively. In spin-integrated mode these resolutions were set to 0.5\dg~and 20 meV. All data were taken at 20 K. The experimental results in the main text were reproduced on fourteen samples with individual annealing preparation over different experimental runs. The magnetic properties were measured using a superconducting quantum interference device in vibrating sample magnetometry mode. Magnetic hysteresis loops were recorded at 5K with the applied field directed perpendicular to the sample plane. The piezo-force microscopy was performed at the NanoXAS endstation at the SLS using plain platinum tip at room temperature. SARPES data from magnetized samples were measured at Soleil CASSIOP{\'E}E beamline.

\subsection*{First principle calculations.}

The \textit{ab-initio} calculations are based on the multiple scattering approach (KKR method) and density functional theory \cite{Minar_RPP}.  Spin-orbit coupling has been naturally included by use of a fully relativistic  four-component scheme. As a first step of our investigations we performed self-consistent calculations  for 3D bulk as well as 2D  semi-infinite surface of \GMT\ within the screened Korringa-Kohn-Rostoker formalism\cite{Minar_RPP}. The corresponding ground state band structures are presented in terms of Bloch spectral functions. The self-consistent results served as an input for our spectroscopic investigations. The ARPES calculations were performed in the framework of the fully  relativistic one-step model of photoemission\cite{JBraun_1SM} in its  spin-density matrix formulation, which accounts properly for the complete spin-polarization vector, in particular for Rashba systems like GeTe. Together with a realistic model for the surface barrier potential, the one-step calculations were decisive to substantiate the \GMT\ spectral features on both qualitative and quantitative  levels. Ground state, as well as one-step calculations of substitutionally disordered \GMT\ has been described by means of coherent potential approximation alloy theory. For simplicity the ground state calculations in Fig.~3b were based on the GeTe lattice structure \cite{Picozzi_AdvM} with added Mn substitutional doping on Ge\nh sites, being the primary reason for the difference in the strength of the Rashba splitting $\alpha$ between theory and experiment reported in Fig.~3.

\section*{Acknowledgemets}


This work was supported by the Swiss National Science Foundation Project No. PP00P2\_144742\/1. Financial support from the German funding agency DFG (SPP1666) and the German ministry BMBF (05K13WMA) is also gratefully acknowledged (H.V., H.E., J.B. and J.M.) G.S. and V.V.V. acknowledge support from the Austrian Science Funds (SFB-025, IRON).
Financial support from the priority program SPP 1666 of the Deutsche Forschungsgemeinschaft (Grant No. EB154/26-1 ) and the German ministry BMBF (05K13WMA) is gratefully acknowledged (H.V.,.H.E.,J.B. and J.M.). J.M. acknowledges CENTEM PLUS (LO1402) project. V.N.S, J.H.D. and G.S initiated and coordinated the project on equal level; H.V. performed the main calculations under supervision of J.M.; supporting calculations were carried out by J.B. and H.E.; SARPES experiments: J.K., S.M., M.F., A.W. and J.H.D.; soft-X ray ARPES experiments: J.K, S.M., F.B. and V.N.S; PFM measurements: N.P.; SQUID measurememnts P.W.; sample growth and structural characterization: V.V.V. and G.S.; data analysis: J.K.; writing of manuscript: J.K., G.S., V.N.S and J.H.D.  All authors extensively discussed the results and the manuscript. The authors declare that they have no competing financial interests. Correspondence and requests for materials should be addressed to G.S. (sample growth and characterization), V.N.S. (SX-ARPES) and J.H.D. (SARPES).


\end{document}